\documentstyle[twocolumn,amssymb,aps,psfig,floats]{revtex}
\begin{document}
\draft

\title{Temperature dependent spatial oscillations in the correlations of
the XXZ spin chain }

\author{Klaus Fabricius
\footnote{e-mail Klaus.Fabricius@theorie.physik.uni-wuppertal.de}}
\address{ Physics Department, University of Wuppertal, 
42097 Wuppertal, Germany}
\author{Andreas Kl{\"u}mper
\footnote{e-mail kluemper@thp.Uni-Koeln.DE}}               
\address{ Institut f{\"u}r Theoretische Physik, Universit{\"a}t zu
K{\"o}ln, Z{\"u}lpicher Str. 77, 50937 K{\"o}ln 41, Germany}
\author{Barry~M.~McCoy
\footnote{e-mail mccoy@insti.physics.sunysb.edu}}               
\address{ Institute for Theoretical Physics, State University of New York,
 Stony Brook,  NY 11794-3840}
\date{\today}
\preprint{ITPSB-98-67}

\maketitle

\begin{abstract}
We study the correlation $<\sigma^z_0\sigma^z_n>$ for the XXZ chain in
the massless attractive (ferromagnetic) region  at positive temperatures
by means of a numerical study of the quantum transfer matrix. We find
that there is a range of temperature where the behavior of the
correlation for large separations is oscillatory with an
incommensurate  period which
depends on temperature. 

\end{abstract}
\pacs{PACS 75.10.Jm, 75.40.Gb}
\section{Introduction}

Recently we \cite{fm}-\cite{fkm} have studied the finite temperature
spin correlation function
 \begin{equation}
S^z(n;T,\Delta)={\rm Tr}\sigma_0^z \sigma_n^z e^{-H/kT}/{\rm Tr}e^{-H/kT}
\label{sz}
\end{equation}
for the spin 1/2 XXZ spin chain \cite{bet}-\cite{yyc}
\begin{equation}
H={1\over 2}\Sigma_{j=0}^{L-1} (\sigma_j^x\sigma_{j+1}^x+
\sigma_j^y\sigma_{j+1}^y+\Delta \sigma_j^z\sigma_{j+1}^z).
\label{ham}
\end{equation}
In \cite{fm} we made a finite size study of $S^z(n;T,\Delta)$ for
chains of size up to $L=18$ and found that for $-1<\Delta<0$ there is
a range of temperature where the correlation function changes sign as
a function of $n$ starting from negative values for $n=1$. For $T$
sufficiently low the correlations for  $-1<\Delta<0$ are negative for
all $n$ whereas for sufficiently large $T$ the correlations are
positive for all $n.$ We referred to this change in sign as a quantum
classical crossover.

In both \cite{fm} and \cite{fkm} we discussed the representation of
this correlation function for the chain with $L\rightarrow \infty$ 
in terms of an expansion in terms of 
eigenvalues and eigenvectors of the quantum 
transfer matrix \cite{suz}-\cite{klu}
\begin{equation}
S^z(n;T,\Delta)=\sum_{j}A_j(\lambda_j/\lambda_0)^{n-1}
\label{expansion}
\end{equation}
where $\lambda_0$ is the maximum eigenvalue of the quantum transfer
matrix and is directly related to the free energy per lattice site by
\begin{equation}
f=-kT\ln \lambda_0+\Delta/2
\end{equation}
This quantum transfer matrix is defined as the limit $N\rightarrow
\infty$ of a matrix of dimension $2^N$ where $N,$ which is even, 
is the ``Trotter
number'' (and is not to be confused with the chain length $L$ which is taken to
infinity.) We have numerically studied
this expansion for values of $N$ as large as 16 and have found several
features of the correlations which have not previously been seen which
substantially expands our  understanding of the phenomena of
quantum-classical cross over. 

From our study we find that as the temperature increases from zero
the following phenomena occur:
\begin{enumerate}
\item There is a temperature $T_V(\Delta)$ below which all amplitudes
$A_j$ are negative and at which temperature many $A_j$ vanish. There
is no other temperature at which any $A_j$ vanishes. For $T<T_V(\Delta)$ all
eigenvalues are real. This temperature vanishes as $1/N$ as
$N\rightarrow \infty.$ We will refer to $T_V(\Delta)$ as the universal
vanishing temperature.

\item There is a temperature $T_C(\Delta)>T_V(\Delta)$ above which
real eigenvalues collide and become complex conjugate pairs. 
At the temperatures where the collision of the eigenvalues occurs the
amplitudes $A_j$ diverge as a square root.
In the
limit $N\rightarrow \infty$ we have $T_C(\Delta)\rightarrow
T_V(\Delta).$

\item There is a temperature $T_L(\Delta)>T_C(\Delta)$ where the
eigenvalue in (\ref{expansion}) with the largest magnitude becomes complex.
As $N\rightarrow \infty$ $T_L(\Delta)$ approaches a nonzero limiting value
from above. We refer to $T_L(\Delta)$ as the lower crossover temperature.

\item There is a temperature $T_U(\Delta)>T_L(\Delta)$  where the
eigenvalue in (\ref{expansion}) with the largest magnitude becomes real
and the corresponding amplitude is positive. At $T_U(\Delta)$ 
eigenvalues with different quantum numbers cross. As $N\rightarrow \infty$
$T_U(\Delta)$ approaches its limiting value from below. We refer to
$T_U(\Delta)$ as the upper crossover temperature.

\end{enumerate}

We conclude that for $T$ between $T_L(\Delta)$ and $T_U(\Delta)$ the
correlation $S^z(n;T,\Delta)$ oscillates and changes sign an infinite 
number of times as $n\rightarrow \infty.$

In the remainder of this letter we will present the evidence to
support and illustrate these conclusions.

\section{The universal vanishing temperature $T_V(\Delta)$}

At $T=0$ it has been known for some time \cite{lp}-\cite{fog} 
when $n\rightarrow \infty$ that for $-1<\Delta<0$
\begin{equation}
S^z(n;0,\Delta)\sim -{1\over \pi^2\theta n^2}+(-1)^n{C(\Delta)\over
n^{1\over \theta}}
\label{zero}
\end{equation}
where $\cos \pi \theta =-\Delta$ and the constant $C(\Delta)$ has
recently been conjectured \cite{luk}. In the very low temperature limit
$T\rightarrow 0$ and $n\rightarrow \infty$ with $nT=r$ fixed where
conformal field theory is applicable \cite{cardy}-\cite{affleck} the large
$n$ behavior of $S^z(n,T,\Delta)$ is obtained from (\ref{zero}) by the
replacement \cite{kbi} $n \rightarrow (\kappa T/2)^{-1}\sinh \kappa r/2$ where
$\kappa=\pi(1-\theta)/ \sin \pi\theta.$ 

From our numerical study we find that for sufficiently low temperatures 
all eigenvalues $\lambda_j$ are real (although not all are positive) and
all amplitudes $A_j$ are negative. We illustrate this in table 1 where
we show the first 13 eigenvalues and amplitudes in (\ref{expansion}) for
$\Delta=-.5, T=.1, N=12.$ We choose $\Delta=-.5$ only for purposes of
illustration. All phenomena discussed in this paper occur generally
for $-1 < \Delta < 0.$
Here, and in the subsequent tables, we give
only the eigenvalues in the subspace which is odd under spin
inversion because the matrix elements in the complementary subspace
are identically zero. We also give the quantum number $k$ which plays
for the quantum transfer matrix the role which momentum plays for
the row transfer matrix. We note for $N\equiv 0 ({\rm mod} 4)$ that
$k=0,\pm 1,\cdots ,\pm (N/4-1),N/4$ and for $N\equiv 2 ({\rm mod} 4)$
that $0,\pm 1,\cdots,\pm ({N-2\over 4}).$ The eigenvalues for $\pm k$
are degenerate because of reflection symmetry of the quantum transfer matrix. 
\begin{table}[here]
\begin{tabular}{|r|r|r|r|r|r|}
   j & $|\lambda_j/\lambda_0|$ &  $\lambda_j$ phase & $|A_j|$ & $A_j$ phase &
  k\\ \hline
  1,2 & .5617564& 0.0 & 2.906227$\times 10^{-2}$& $\pi$& $\pm 1$ \\     
  3 & .3954749& $\pi$&  5.619120$\times 10^{-3}$& $\pi$&   0 \\    
  4,5 & .3346469& 0.0&  3.258793$\times 10^{-2}$& $\pi$&  $\pm 2$ \\    
  6 & .2323767& 0.0&  4.927463$\times 10^{-2}$& $\pi$&   3 \\    
  7,8 & .1911818& 0.0&  1.733959$\times 10^{-2}$& $\pi$&  $\pm 2$ \\        
  9,10 & .1833028& $\pi$ & 4.491119$\times 10^{-3}$&$\pi$& $\pm 1$ \\        
 11,12 & .1764959& 0.0&  1.292250$\times 10^{-2}$& $\pi$ &$\pm  1$ \\     
 13 & .1728539& 0.0&  1.158930$\times 10^{-2}$& $\pi$ &  0 \\ 
 \end{tabular}   
\caption{Eigenvalues below $T_V(\Delta).$
The first 13 eigenvalues $\lambda_j$ and
amplitudes $A_j$ for $\Delta=-.5, T=.10$ and $N=12.$ The value of
$\lambda_0$ is 26.67983.}
\end{table}  
  
As we increase the temperature the eigenvalues remain real and the
amplitudes all remain negative until we reach a temperature
$T_V(\Delta)$ where a most dramatic effect occurs. Namely vast numbers
of the amplitudes all vanish and the eigenvalues with vanishing
amplitudes occur in degenerate sets in which all allowed values of $k$
occur for $N\equiv 2 ({\rm mod} 4)$. (For $N\equiv 0 ({\rm mod} 4)$ the
eigenvalue with $k=N/4$ is degenerate but the amplitude does not
vanish). As $N\rightarrow \infty$ there are an infinite number of
these degenerate sets. For
$\Delta=-.5$ and $N=12$ we find that $T_V(-.5)$ is slightly below 
$.137832224$. In table 2 we give the first 13 eigenvalues and
amplitudes for this case. Because the temperature is slightly above
the universal vanishing temperature the sign of the amplitudes with
the vanishingly small magnitudes are positive. The order of the
eigenvalues 6-11 in table 2 is correct even though to the number of
significant figures given the eigenvalues are identical.
It should be noted that
in table 2 all eigenvalues are still real. 
\begin{table}[here]
\begin{tabular}{|r|r|r|r|r|r|r|}
  $j$&$j_1$&$|\lambda_j/\lambda_0|$ &$\lambda_j$ phase& $|A_j|$& $A_j$ phase&
  k  \\ \hline
  1,2&1,2&  .4791320 & 0.0&  4.285003$\times 10^{-2}$& $\pi$ &$\pm 1$ \\     
  3&3&  .3177245 & $\pi$&  7.130329$\times 10^{-3}$ &$\pi$ &  0 \\   
  4,5&4,5&  .2424384& 0.0&  5.180946$\times 10^{-2}$& $\pi$ &$\pm  2$ \\    
  6,7&7,8&  .1737014&0.0 & 1.726434$\times 10^{-10}$&  0.0&$\pm  2$ \\   
  8,9&9,10&  .1737014& 0.0&  5.754777$\times 10^{-11}$&  0.0 &$\pm  1$ \\     
 10&13& .1737014& 0.0&  4.316086$\times 10^{-11}$&  0.0 &  0 \\   
 11&6&  .1737014&  0.0&  5.186948$\times  10^{-2}$& $\pi$ &  3 \\   
 12,13&9,10&  .1328673&  $\pi$&4.055662$\times 10^{-3}$& $\pi$ &$\pm  1$ \\   
   \end{tabular}
\caption{Eigenvalues at $T_V(\Delta).$
The first 13 eigenvalues $\lambda_j$ and amplitudes $A_j$
for $\Delta=-.5, T= 0.137832224$ and $N=12.$ The value of $\lambda_0$
is 10.7928. The column $j_1$ is the value of $j$ of the
corresponding eigenvalue for $T=.1$ in table 1. }
\end{table}

The numerically determined values of $T_V(-.5)$ for $N=10,12,14,16$  are
given in table 6. The $N$ dependence is extremely well fit for $6\leq
N \leq 16$ by
$T_V(-.5;N)=1.6539/N$ and therefore  $T_V(\Delta)$ vanishes as
$N\rightarrow \infty.$ We thus conclude that the data for
$T<T_V(\Delta)$ and $N$ finite is not relevant to the genuine quantum
system with $N\rightarrow \infty.$

\section{Complex eigenvalues}

As the temperature is increased above the universal vanishing
temperature $T_V(\Delta)$ we very soon reach
a temperature where two real eigenvalues collide
and become complex conjugate pairs. This happens first with
eigenvalues which are very small. 
As $T$ increases beyond the temperature of first collision $T_C(\Delta)$
successively larger eigenvalues collide. At each collision the
amplitudes diverge however the correlation functions remain finite due
to a perfect phase difference of $\pi$ at the point of level
collision. At the temperature of collision the relevant exponential
decay terms in (\ref{expansion}) are replaced by $An(\lambda_j/\lambda_0)^{n-1}.$ 

 We illustrate the collision of eigenvalues in table 3 where for
$\Delta=-.5$ and $N=12$ we give the first 13 eigenvalues and
amplitudes at $T=.151493876679$ which is slightly above the collision
temperature for the pairs of eigenvalues 4,5 and 6,7. 
Because the temperature is above collision the
eigenvalues and amplitudes of the eigenvalues 4,5,6,7 are complex
and all four combinations of the signs of these two phases are to be used. 
It is clear from the table that
the amplitudes are diverging and the change in phase of the amplitude
from real to $\pi/2$ suggests that this divergence must be a
square root. Further analysis of the behavior of the amplitudes in the
vicinity of the collision shows that this is indeed the correct behavior.
\begin{table}[here]
\begin{tabular}{|r|r|r|r|r|r|}
  j&$|\lambda_j/\lambda_0|$ &$\lambda_j$ phase& $|A_j|$& $A_j$ phase&  k  \\ \hline
  1,2&.4481334& 0.0&  4.941102$\times 10^{-2}$&  $\pi$ &$\pm  1$\\      
  3&.2918946& $\pi$&  7.509636$\times 10^{-3}$ &$\pi$  & 0\\    
   4-7&.1909054& $\pm 1.1899\times 10^{-6}$&4.245$\times 10^{+3}$ 
  & $\pm 1.5708$ & $\pm 2$\\        
 8.9& .1746361 &0.0 &5.837584$\times 10^{-3}$ &0.0&$\pm 1$\\
 10&.1741270&  0.0&  4.018312$\times 10^{-3}$ &0.0  & 0\\    
 11&.1561094&  0.0& 5.198129$\times 10^{-2}$ & $\pi$ &  3\\    
 12,13&.1180399& $\pi$ & 3.785983$\times 10^{-3}$& $\pi$ & $\pm 1$ \\       
  \end{tabular}

\caption{A collision of eigenvalues below $T_L(\Delta).$
 The first 13 eigenvalues $\lambda_j$ and amplitudes $A_j$
for $\Delta=-.5, T= 0.151493876679$ and $N=12.$ The value of $\lambda_0$
is 8.224613. The order of eigenvalues is the same as in table 2.
The phases and the magnitude of
 $|A_j|$  for eigenvalues 4-7 have only been
 determined to four places because of the numerical instabilities
 resulting from the divergence of $A_j.$ }
\end{table}

\section{The onset of oscillations at the lower crossover 
temperature  $T_L(\Delta)$}

As the temperature is increased further we reach a temperature
$T_L(\Delta)$ where the two eigenvalues with the largest magnitudes
collide and produce a complex conjugate pair. These eigenvalues are in
the sectors $k=\pm 1.$ For $\Delta=-.5$ and
$N=12$ this occurs slightly below  $T=.22782901.$ 
In table 4 we give the largest 13 eigenvalues at this temperature.
\begin{table}[here]
\begin{tabular}{|r|r|r|r|r|r|}
  j&$|\lambda_j/\lambda_0|$ &$\lambda_j$ phase& $|A_j|$& $A_j$ phase
   &  k  \\ \hline
  1-4&.2388172&  $\pm $1.155$\times 10^{-4}$&  1.9078$\times 10^{+2}$
& $\pm 1.570925$&$\pm 1$ \\  
  5&.1815640&     $\pi$    &  7.424416$\times 10^{-3}$& $\pi$  &   0 \\   
  6&.1779166&  0.0         &  2.511491$\times 10^{-2}$& 0.0  &   0 \\   
  7-10&.1347183& $\pm.4576581$ & 2.968286$\times 10^{-2}$ 
  & $\pm2.561349$&$\pm 2$ \\   
 11&.0903745&  0.0         &  4.730570$\times 10^{-2}$&  $\pi$   &   3 \\
 12,13&.0641918&  $\pi$ &  2.135921$\times 10^{-3}$&  
  $\pi$   & $\pm  1$ \\  
 \end{tabular}
\caption{Eigenvalues at $T_L(\Delta).$  
The first 13 eigenvalues $\lambda_j$ and amplitudes $A_j$
for $\Delta=-.5, T= 0.22782901.$ and $N=12.$ The value of $\lambda_0$
is  4.173698.}
\end{table}
The dependence of $T_L(-.5)$ on $N$ is given in table 6.
This $N$ dependence is well fit by
$T_L(-.5;N)=.2015+3.76/N^2$ and this has been used to extrapolate to
$N\rightarrow\infty.$

At the lower crossover temperature the leading asymptotic
behavior of the correlation is
\begin{equation}
S^z(n;T_L(\Delta),\Delta)\sim A n(\lambda/\lambda_0)^{n-1}.
\end{equation}
As the temperature increases further the leading behavior of the
correlation is oscillatory with a wavelength which decreases as T increases.

\section{The completion of the crossover at the upper crossover
temperature $T_U(\Delta)$}

In tables 2,3 and 4 we note that there is a positive  
eigenvalue with $k=0$  
which has a positive amplitude. At the lower crossover temperature
this eigenvalue does not have the largest magnitude and has $j=6$ in
table 4  lying below
the complex conjugate pairs 1,2 and 3,4 and the negative
eigenvalue with the negative amplitude at j=5. However, as $T$
is increased from $T_L(\Delta)$ the magnitude of this eigenvalue
increases relative to the magnitude of the leading complex conjugate
eigenvalue and at a temperature $T_U(\Delta)$ the magnitudes of these
eigenvalues cross. For $\Delta=-.5$ and $N=12$ this occurs very near
T=.3611. In table 5 we give the first 13 
eigenvalues at a slightly higher temperature where the real eigenvalue
has the largest magnitude. 
Such a crossing without collision is possible
because the eigenvalues lie in different sectors $k.$
We call the temperature where this crossing occurs the upper
crossover temperature. The dependence of $T_U(-.5)$ on $N$ is given
in table 6. This $N$ dependence is well fit by $T_U(-.5;N)=
.4328-.86/N$ and this
has been used to extrapolate to $N\rightarrow \infty.$

\begin{table}[here]
\begin{tabular}{|r|r|r|r|r|r|r|}
  j&$j_4$&$|\lambda_j/\lambda_0|$ &$\lambda_j$ phase& $|A_j|$& $A_j$ phase&  k  \\ \hline
  1&6  &.1862180&  0.0         &  6.113537e-02&  0.0         &   0 \\    
  2-5&1-4  &.1858948& $\pm.5968364$&  5.134231e-02& $\pm2.118513$& $\pm 1$ \\  
  6-9&7-10  &.0863447& $\pm .7691693$&  2.144874e-02&$\pm  2.723044$&
  $\pm 2$ \\   
   10&5  &.0850952&   $\pi$      &  3.715138e-03&    $\pi$     &   0 \\   
 11&11 &.0428430&  0.0         &  3.407129e-02&    $\pi$     &   3 \\   
 12-13&12-13 &.0275197&   $\pi$&  6.337585e-04& $\pi$& $\pm 1$   \\ 
 \end{tabular}
\caption{Eigenvalues at $T_U(\Delta).$ 
The first 13 eigenvalues $\lambda_j$ and amplitudes $A_j$
for $\Delta=-.5, T=.3620 $ and $N=12.$ The value of $\lambda_0$
is  2.679573. The column $j_4$ is the value of $j$ of the
corresponding eigenvalue in table 4.}

\end{table}
For all temperatures above the upper crossover  temperature the leading
eigenvalue is positive and has a positive amplitude. Consequently for
large $n$ the correlation will be monotonic and positive.

\begin{table}[here]
\begin{tabular}{|r|r|r|r|}
  N&$T_V$&$T_L$ &$T_U$  \\ \hline
  10&0.1653900  &0.2391  & 0.3463  \\
  12&0.1378322  &0.2278  & 0.3611  \\
  14&0.1181419  &0.2207  & 0.3716   \\
  16&0.1033742  &0.2162  & 0.3791  \\  
  $\infty$&0.0  &  0.2015& 0.4328
\end{tabular}
\caption{The numerically determined values of the temperatures
$T_V(\Delta),T_L(\Delta)$ and $T_U(\Delta)$ with $\Delta=-.5$ for
various values of $N.$} 
\end{table}
\section{Conclusions}

In \cite{fm} we found that there is a range of temperatures in which
the correlation function $S^z(n;T,\Delta)$ changes sign from negative
to positive as $n$ is increased and in \cite{fkm} we discussed level
crossing as a mechanism for this effect. The studies of the
eigenvalues of the quantum transfer matrix presented here demonstrate
that instead of the single zero seen in the previous finite size
studies there is a range of temperatures from $T_L(\Delta)$ to
$T_U(\Delta)$ where the correlation function has an oscillatory
behavior with a temperature dependent wavelength. This phenomenon has
not been seen before in any study of the $XXZ$ chain and can only
occur because the non hermitian quantum transfer matrix  can have
complex eigenvalues. These eigenvalues can be studied analytically by
using the Bethe-like equations derived in \cite{klu}. However,
the necessity of considering complex eigenvalues requires
considerations which have not been encountered in previous studies of
related Bethe Ansatz equations. As stated previously \cite{fkm}
we believe that the quantum-classical crossover which is
seen in the quantum transfer matrix is deeply connected with the
bound states at $T=0$ seen in the spectrum of the Hamiltonian (\ref{ham}).    
Analytic investigations of these questions will be pursued elsewhere.

{\bf Acknowledgments}

This work is supported in part by the National Science Foundation
under Grant No. DMR-03543.
A.K. acknowledges  financial   support  by the   {\it  Deutsche
Forschungsgemeinschaft} under grant  No.   Kl~645/3
and support by the research program of the 
Sonderforschungsbereich 341, K\"oln-Aachen-J\"ulich.

\end{document}